\documentclass[conference]{IEEEtran}
\IEEEoverridecommandlockouts
\usepackage{cite}
\usepackage{amsmath,amssymb,amsfonts}
\usepackage{algorithm}
\usepackage{algpseudocode}
\usepackage{graphicx}
\usepackage{textcomp}
\usepackage{xcolor}
\usepackage{balance}
\usepackage{epsfig}
\usepackage[left=0.6in,right=0.6in,top=0.72in, bottom=0.72in]{geometry}
\usepackage{float}
\usepackage{verbatim}
\usepackage{tabularx}
\usepackage{svg}

\def\BibTeX{{\rm B\kern-.05em{\sc i\kern-.025em b}\kern-.08em
    T\kern-.1667em\lower.7ex\hbox{E}\kern-.125emX}}

\begin{document}

\title{Text2Net: Transforming Plain-text To A Dynamic Interactive Network Simulation Environment}

\author{
\IEEEauthorblockN{Alireza Marefat, Abbaas Alif Mohamed Nishar, Ashwin Ashok}\\
\IEEEauthorblockA{\textit{[amarefatvayghani1, amohamednishar1]}@student.gsu.edu,} \textit{aashok@gsu.edu}\\
\\ \IEEEauthorblockA{\textit{Georgia State University, Atlanta, USA}}
}

\maketitle
\noindent{\footnotesize \textcopyright\ 2025 IEEE. Personal use of this material is permitted. Permission from IEEE must be obtained for all other uses, \\
in any current or future media, including reprinting/republishing this material for advertising or promotional purposes, \\
creating new collective works, for resale or redistribution to servers or lists, or reuse of any copyrighted component of this work in other works.}

\begin{abstract}
This paper introduces Text2Net, an innovative text-based network simulation engine that leverages natural language processing (NLP) and large language models (LLMs) to transform plain-text descriptions of network topologies into dynamic, interactive simulations. Text2Net simplifies the process of configuring network simulations, eliminating the need for users to master vendor-specific syntaxes or navigate complex graphical interfaces. Through qualitative and quantitative evaluations, we demonstrate Text2Net's ability to significantly reduce the time and effort required to deploy network scenarios compared to traditional simulators like EVE-NG. By automating repetitive tasks and enabling intuitive interaction, Text2Net enhances accessibility for students, educators, and professionals. The system facilitates hands-on learning experiences for students that bridge the gap between theoretical knowledge and practical application. The results showcase its scalability across various network complexities, marking a significant step toward revolutionizing network education and professional use cases, such as proof-of-concept testing.
\end{abstract}


\section{Introduction}

Network simulators and emulators are essential tools in computer science (CS) education, allowing students to explore and experiment with complex network behaviors without relying on physical hardware. These tools are also widely used in industry for testing and validation purposes. Simulators are software engines that replicate various networking scenarios to test protocols, configurations, and network dynamics. Cisco Packet Tracer \cite{janitor2010visual} is a popular simulator for beginners \cite{allison2022simulation}, while GNS3 \cite{neumann2015book} caters to advanced users with its capability to simulate real device images. However, these simulators primarily model network behavior and may not fully replicate real-world dynamics.
In contrast, emulators like EVE-NG~\cite{EVE-NG} offer a more realistic solution by supporting actual device system code images (e.g., Cisco IOS), enabling accurate emulation of real-world operations. EVE-NG’s robust features make it particularly valuable for advanced education and professional training, allowing users to engage with complex network topologies in realistic environments \cite{sharma2024comparison}.


\vspace{1mm}\noindent{\bf Challenges with network simulators in education.} Despite advancements in simulation tools, significant barriers hinder their effective use in education. Traditional tools often require mastering complex, vendor-specific command syntax, making setup processes repetitive and time-consuming~\cite{sierszen2017teaching}. This focus on memorizing configurations detracts from understanding core concepts and designing network architectures, which are far more valuable skills. Furthermore, the wide variety of simulation tools introduces challenges such as poor maintenance, the dilemma of paid versus open-source options, and difficulties in transferring experiments between platforms~\cite{marquardson2019simulation}.

\vspace{1mm}\noindent{\bf Text2Net: Bridging the gap using plain text and AI.}
Text2Net provides an innovative solution by enabling users to create and interact with network simulations using plain-text English instead of vendor-specific syntax. Leveraging advancements in natural language processing (NLP) and large language models (LLMs), Text2Net interprets user inputs and translates them into actionable simulation commands. While LLMs are powerful, they are prone to issues such as errors and inaccuracies. Text2Net addresses these challenges by eliminating the need for technical expertise, simplifying the simulation process, and shifting the focus from command syntax to conceptual learning. This approach enhances accessibility and efficiency, reducing the effort required to deploy and manage network scenarios.

This paper focuses on the development of the Text2Net engine and its application in network education. We present the system architecture and demonstrate its usability through a case study involving the EVE-NG tool. A complete prototype implementation is evaluated qualitatively through user surveys and quantitatively by comparing task completion steps and time between Text2Net and EVE-NG.

\section{Related Works}
The intersection of AI and network management has prompted several innovative approaches, each aimed at enhancing the adaptability and efficiency of network systems. 
NetGPT \cite{chen2024netgpt} has been developed as an AI-native network architecture that strategically deploys LLMs both at the edge and cloud to optimize personalization and efficiency. The architecture highlights improvements in network management and user intent inference by integrating communications and computing resources more deeply \cite{tong2023ten}. Similarly, NetLM \cite{wang2023network} introduces an AI-driven architecture to enhance autonomous capabilities in network management, notably in complex 6G environments. The system leverages multi-modal representation learning to integrate diverse network data, aiming to refine network intents and autonomously manage network operations.
ABC (Automatic Bottom-up Construction) \cite{ding2023abc} revolutionizes the configuration knowledge base for multi-vendor networks by automating the alignment and generation of configuration templates through natural language processing and active learning, significantly reducing the manual effort typically required.
CONFPILOT \cite{zhao2023confpilot} employs a retrieval-augmented generation framework to translate natural language intents into precise network configuration commands. This system not only accelerates configuration processes but also enhances accuracy with its innovative use of a retrained BERT model and a parameter description-enhanced BM25 algorithm, which together improve the retrieval and matching of network commands.
NetCR \cite{guo2023netcr} utilizes a knowledge graph to facilitate manual network configurations, providing adaptive recommendations that enhance the efficiency and accuracy of network operations across various devices. This tool underscores the potential of using structured knowledge to streamline network management tasks in multi-vendor environments. To the best of our knowledge, Text2Net is the first initiative that directly integrates AI, specifically NLP, into network simulation for educational purposes and beyond. While prior works have explored the use of AI to enhance network management and configuration, Text2Net uniquely applies these technologies to simplify and democratize the learning and execution processes in network simulations. 
\begin{figure}[t!]
    \centering
    \includegraphics[width=\columnwidth]{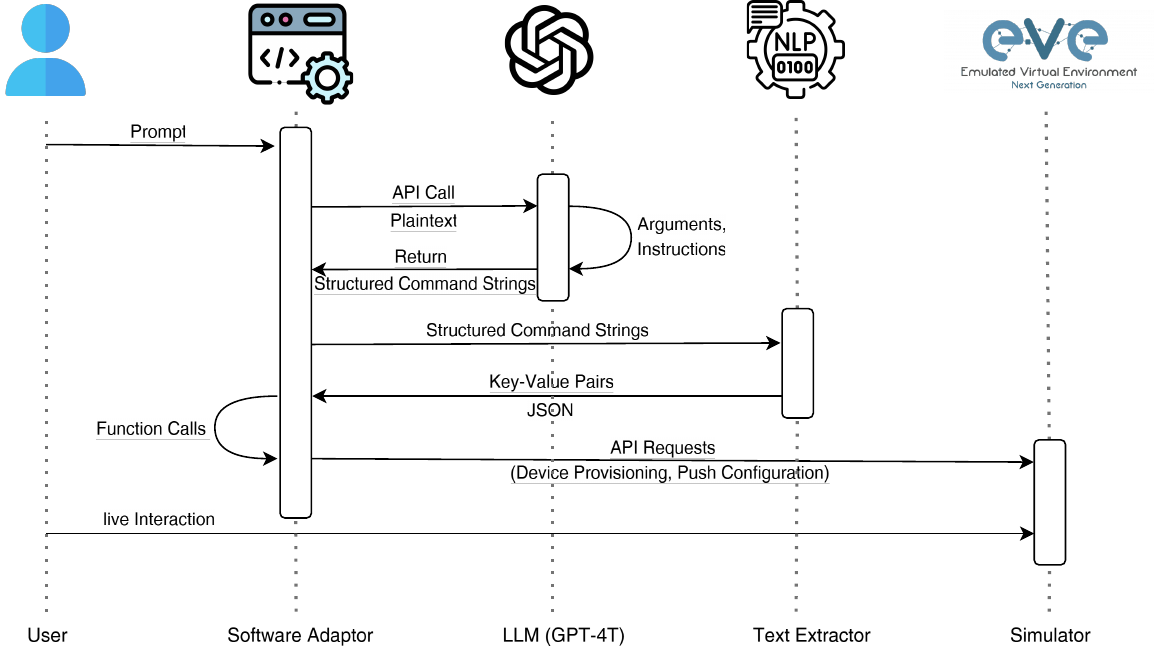}
    \caption{Text2Net system model and pipeline}
    \vspace{-4mm}
    \label{fig: system model}
\end{figure}
\section{Text2Net Methodology}
As depicted in Figure~\ref{fig: system model}, Text2Net comprises five modules: User, Software-Adaptor, instructed LLM, Text Extractor, and Simulator. The user inputs a network topology scenario, which the Software-Adaptor forwards to the instructed LLM, OpenAI's ChatGPT-4T, via API calls. The LLM processes this input and returns Structured Command Strings (SCS) to the Software-Adaptor. Utilizing NLP tools (SpaCy), along with RegEx and pattern matching, the Software-Adaptor extracts the desired key-value pairs, formatting them into a JSON dictionary. This JSON file is input to the EVE-NG emulator to provision the live network topology and configurations. 

\begin{figure}[ht!]
    \centering
    \includegraphics[width=\columnwidth]{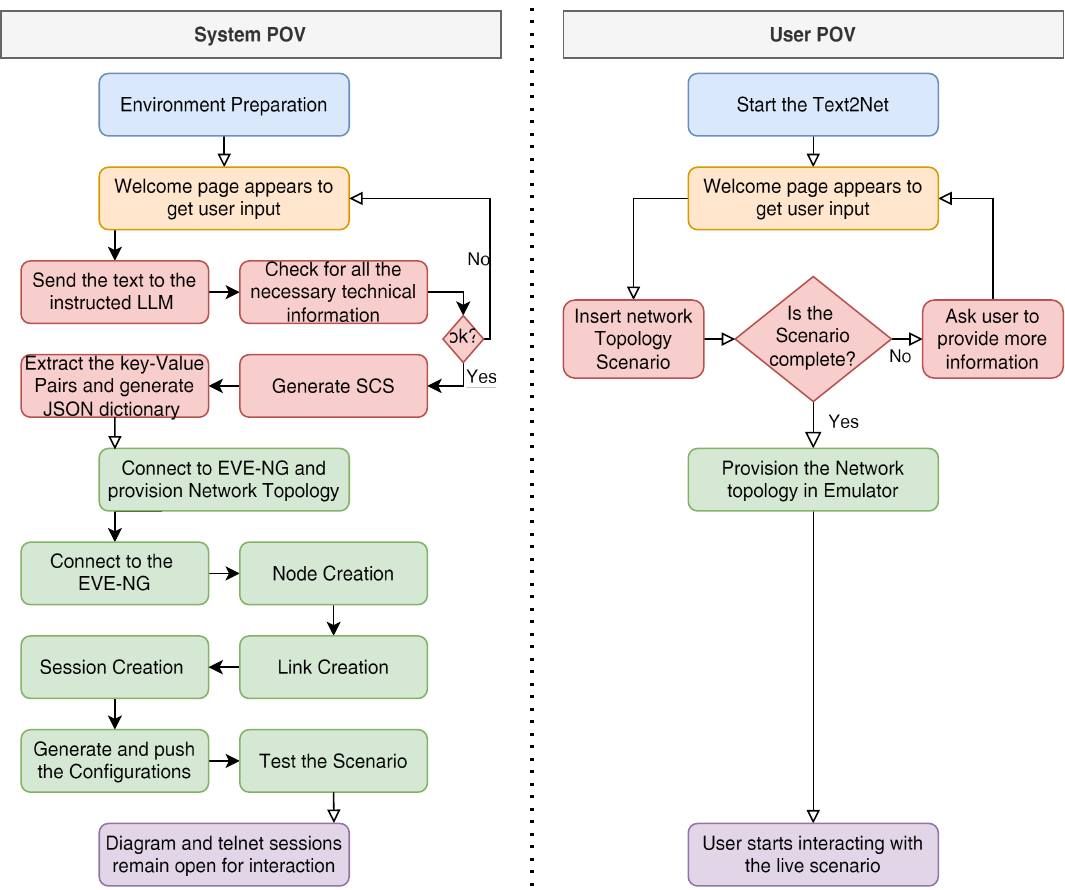}
    \caption{System flow perspectives.}
    \vspace{-3mm}
    \label{fig: flowchart}
    \vspace{-3mm}
\end{figure}

\noindent{\bf Overall System Flow.}
Figure ~\ref{fig: flowchart} depicts the system model, showcasing both, the system and user perspectives.
As illustrated on the right side of Figure~\ref{fig: flowchart}, the system flow from the user perspective begins with the initiation of Text2Net followed by the display of the welcome page. Text2Net greets the user and prompts for a network topology scenario. If the input scenario is valid and includes all required technical details, the system proceeds to provision the network topology in the emulator. If the scenario lacks necessary information, the system requests further details from the user. Ultimately, the user can interact with the fully configured live network topology.
The left side of Figure~\ref{fig: flowchart} illustrates the background processes of the system, which operate transparently to the user and concurrently with the functions depicted on the right side, highlighted in the same color, which will be discussed in detail in this section. 

\vspace{-2mm}
\subsection{System preparation}

The initial setup of Text2Net involves using EVE-NG, a network emulator (it is beyond simulator that replicates real-world environments and supports actual device images from manufacturers like Cisco, Juniper, and HPE). For Text2Net, Cisco devices are primarily used due to their commonality in networking. EVE-NG can be deployed either via an ISO file on virtual machines like VMware or directly on physical hardware to avoid performance issues associated with nested virtualization. To broaden Text2Net’s accessibility, it is hosted on the Google Cloud Platform (GCP) using an n2-standard-8 machine with 8 vCPUs and 32 GB of memory, running Linux Ubuntu. After the EVE-NG installation, the system is configured with a static IP, and HTTPS and SSH ports are opened to ensure it is accessible from any location, verified by navigating to the public IP address in a web browser to reach the EVE-NG login screen.

To leverage OpenAI's ChatGPT-4T for Text2Net, we trained the model to interpret and generate SCS from plain text descriptions of network topologies, commonly presented in computer network lectures. The model was trained to precisely extract and structure key information into command strings with key-value pairs essential for network topology provisioning.
The model recognizes detailed textual descriptions of network setups, outputting accurate command strings without superfluous content. For valid, complete inputs, the model confirms with returning the phrase "Understood", moving to the next phase. For inputs that are empty, incomplete, or incorrect, it prompts the user to refine their input.
The initial user interaction with Text2Net involves a user-friendly interface where users are prompted to input network topologies in plain English. This input becomes the basis for generating network configurations. A significant challenge was standardizing how users describe network topologies; this was addressed through a qualitative survey to establish a standard input format.
Text2Net is equipped to assess the validity and completeness of user inputs, ensuring no essential details, like IP formatting or technical configurations, are missing. This capability ensures the system efficiently transitions from user input to network configuration. 
When an input scenario includes an invalid IP address, such as "192.168.0.300," Text2Net automatically detects the error. For configurations involving specific protocols like static routing that lack necessary details, the system does not simply accept the input. Instead, it prompts the user \textit{``Please provide additional details about the static route"} before proceeding with generating the SCS.

\vspace{-2mm}
\subsection{Extracting Structured Command Strings}
To efficiently extract key-value pairs from the plain text, the text is decomposed into segments known as Structured Command Strings a.k.a. SCSs. These SCSs are derived from the plain text by employing the instructed GPT-4T model. Each SCS consists of short strings that encapsulate one or a few specific key-value pairs, ensuring clarity and specificity in data extraction.
Thanks to prompt engineering, the system is able to extract the same SCSs as an output from the following different scenarios that explain the same network topology with different styles of explanation. Figure~\ref{fig: network topology} shows the network diagram for which we have considered input/output configuration testing across three different user-input scenarios that may be possible:

\begin{figure}[t!]
    \centering
    \includegraphics[scale=0.3]{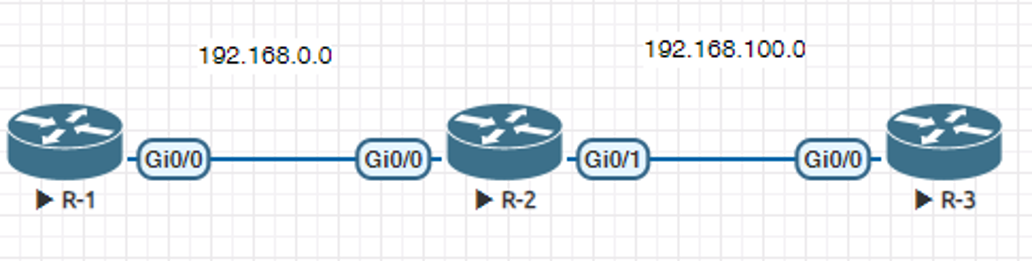}
    \caption{Static-route Scenario Network Topology}
    \vspace{-6mm}
    \label{fig: network topology}
\end{figure}

\vspace{1mm}\noindent{\bf Scenario 1 -}
\textit{“R-1” is a router that is connected to “R-2”. “R-1” interface gi 0/0 has IP address 192.168.0.1/24 and is connected to  “R-2” interface Gi 0/0 with IP address 192.168.0.2/24. 
R-2 is connected to “R-3” via interface Gi 0/1 using IP address 192.168.100.1/24. “R-3” is connected back to “R-2” using interface Gi 0/0 with IP address 192.168.100.2/24. 
A static route is configured on “R-1” to reach “R-3” as well as a static router on “R-3” to reach to “R-1” through “R-2”.}

\vspace{1mm}\noindent{\bf Scenario 2 -}
\textit{The network has three routers: R-1, R-2, and R-3, interconnected in a specific manner.
R-1 connects to R-2 through its interface Gi 0/0, with the IP address 192.168.0.1/24, while R-2's corresponding interface, Gi 0/0, has the IP address 192.168.0.2/24.
R-2 establishes a connection with R-3 via interface Gi 0/1, with R-2 assigned the IP address 192.168.100.1/24 for this link.
The reverse connection from R-3 to R-2 is achieved through R-3's interface Gi 0/0, configured with the IP address 192.168.100.2/24.
For seamless communication between R-1 and R-3, static routes are set up on both routers through R-2, ensuring efficient routing between them.}

\begin{figure}[ht!]
    \centering
    \includegraphics[width=1\columnwidth]{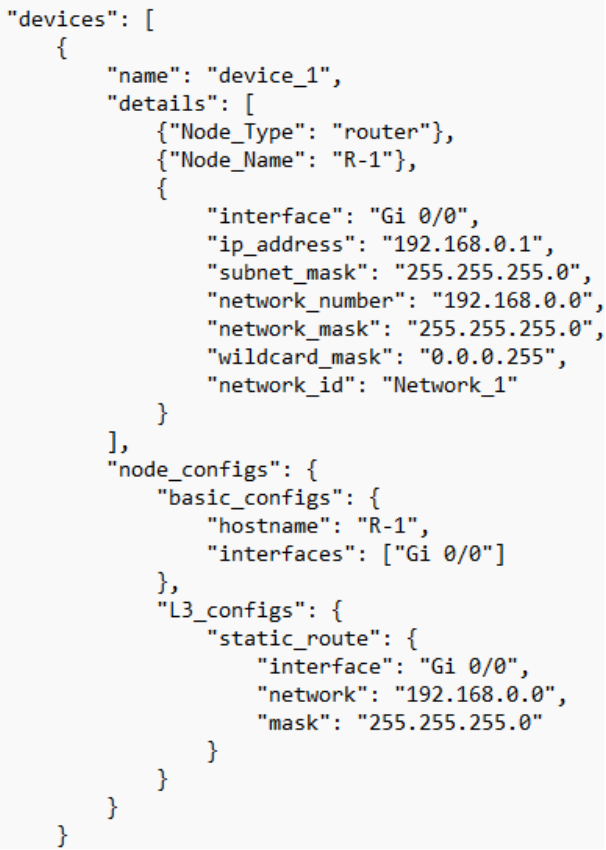}
    \caption{Structured Command Strings (SCSs)}
    \vspace{-4mm}
    \label{fig: SCS}
\end{figure}


Figure~\ref{fig: SCS} shows the same output as SCS from the two above scenarios in our system. Such examples demonstrate Text2Net's capability to deliver consistent output across different storytelling (variable network configuration explanations from users) approaches, provided that the underlying network topology remains the same. This highlights the robustness of this system in recognizing and interpreting the essential elements of network configurations, even when the narrative descriptions vary. This consistency ensures that Text2Net can be reliably used, and is easy to program for changes in input styles, in educational settings where diverse narrative styles are employed to describe similar network setups.

The scenario below illustrates how the same network topology can be described differently, akin to the variations tested in the previous two scenarios. However, there is a key difference, this description lacks detailed information about ``static routing". This incomplete scenario provides an opportunity to observe how Text2Net manages scenarios where critical information is missing.

\begin{small}
\begin{algorithm}[t]
\caption{Text2Net Data Extraction and Structuring}
\label{algorithm}
\begin{algorithmic}[1]  
\State \textbf{Input}: $\mathcal{D} = \{d_1, d_2, \ldots, d_n\}$ 
\State \textbf{Output}: $\mathcal{J}$ 
\State Initialize $\mathcal{J} = \{\texttt{devices}: \{\}, \texttt{connections}: \{\}\}$
\State Set $c = 1$ and $\mathcal{I}$ as an empty map.
\For{each $(k, v) \in \mathcal{D}$}
    \If{$k$ has no comma}
        \State Initialize $\mathcal{V}, \mathcal{C} = \{\}, \mathcal{L} = \{\}$
        \For{each $l \in v$}
            \State $l \gets$ TrimSpaces(RemoveKey($l, k$))
            \If{"type" in $l$} \State $\mathcal{V}$.details.append(\{Node\_Type: ExtractNode($l$)\})
            \ElsIf{"name" in $l$} \State $\mathcal{V}$.details.append(\{Node\_Name: ExtractName($l$)\})
            \EndIf
            \If{"interface" in $l$}
                \State $iface \gets$ ExtractInterfaceDetails($l$)
                \State Update network attributes in $iface$ 
                \State AssignUniqueID($iface, \mathcal{I}, c$)
                \State $\mathcal{V}$.details.append($iface$)
            \EndIf
        \EndFor
        \State $\mathcal{C}$.hostname $\gets$ FindHostname($\mathcal{V}$.details)
        \State $\mathcal{C}$.interfaces $\gets$ FindInterfaces($\mathcal{V}$.details)
        \State $\mathcal{L}$.static\_route $\gets$ FindStaticRoutes($\mathcal{V}$.details)
        \State $\mathcal{V}$.node\_configs $\gets$ \{basic: $\mathcal{C}$, L3: $\mathcal{L}$\}
        \State $\mathcal{J}$.devices.append($\mathcal{V}$)
    \Else
        \State $con \gets$ ParseConnectionDetails($v$)
        \State $\mathcal{J}$.connections.append($con$)
    \EndIf
\EndFor
\State \textbf{Return} $\mathcal{J}$
\end{algorithmic}
\end{algorithm}
\end{small}

\noindent{\bf Scenario 3-}
\textit{This network architecture is designed to facilitate efficient communication between multiple network segments, each identified by distinct IP subnets.
Routers R-1, R-2, and R-3 serve as intermediaries for routing data packets between these segments.
R-1 functions as the gateway router, connecting a potential local network segment to the wider network. It is directly linked to R-2 through its interface Gi 0/0, with R-1's IP address on this interface being 192.168.0.1/24 and R-2's IP address set to 192.168.0.2/24.
R-2 operates as a central hub, facilitating connections between multiple network segments. It interfaces with R-1 through Gi 0/0 and with R-3 through Gi 0/1. On interface gi 0/1, R-2 is assigned the IP address 192.168.100.1/24.
R-3 serves as a bridge between different network segments. It connects back to R-2 through its interface Gi 0/0, configured with the IP address 192.168.100.2/24.
The network's functionality relies on the careful configuration of IP addresses and static routes. This ensures that data packets are routed efficiently between devices connected to R-1, R-2, and R-3, facilitating seamless communication across the entire network infrastructure.}

In this case, the SCSs is generated but not for the static route section. Text2Net detected the missing information and specifically asked about it when returned ``However, I need additional information about the static routing configuration to provide complete command strings. Could you specify the source, destination, and through devices for each static route?''

\vspace{-2mm}
\subsection{Extracting Key-value pairs}

To develop a comprehensive key-value pair dictionary, the first step is to establish a detailed entity relationship. Understanding the scope of the system is crucial for designing the relationships between entities to structure the corresponding JSON dictionary effectively. In the current phase of Text2Net, we leveraged RegEX and pattern matching to implement the system's functionality for routing in networking. However, scaling this approach to cover all networking concepts would be labor-intensive and inefficient. As future work, we aim to explore more on NLP techniques, also including Retrieval Augmented Generators (RAGs) in our model, to enhance scalability and extend the system's capabilities.


\begin{figure}[ht!]
    \centering
    \includegraphics[width=0.55\columnwidth]{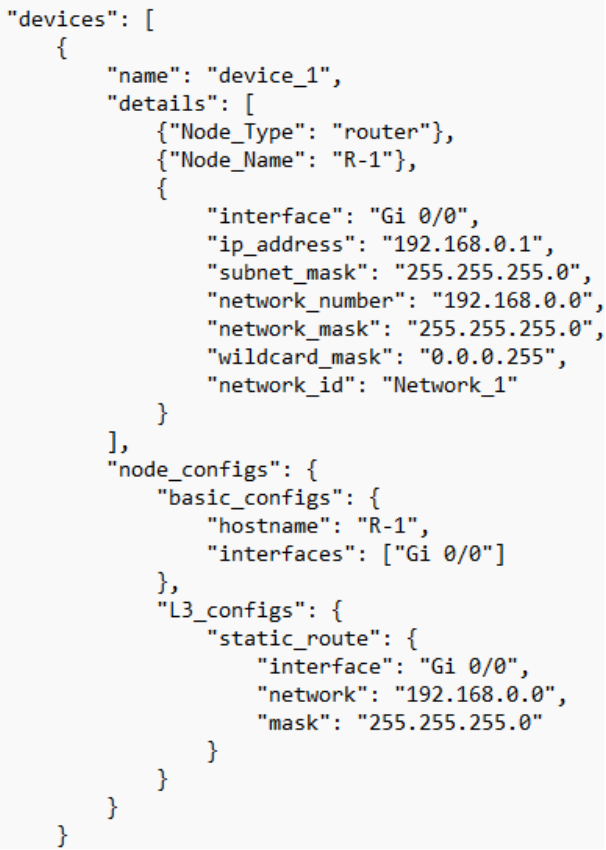}
    \caption{key-value pairs output}
    \vspace{-3mm}
    \label{fig: key-value pairs output}
\end{figure}

Algorithm~\ref{algorithm} facilitates the structured extraction and processing of network topology data. $\mathcal{D} = \{d_1, d_2, \ldots, d_n\}$ represents the set of all devices, where in  $\mathcal \{d_1, d_2, \ldots, d_n\}$ each element is a tuple containing key-value pairs $\mathcal(k, v)$ that describe network devices parameters and their information. The output, $\mathcal{J}$, is a JSON object structured to include detailed device and connection configurations necessary to be used for the network simulation. During processing, $\mathcal{V}$ serves as a temporary dictionary to accumulate the detailed attributes of each device, while $\mathcal{C}$ and $\mathcal{L}$ store basic and Layer 3 configurations respectively. \textit{l} represents each line of SCS that the algorithm iterates through for extraction. Functions such as \verb|ExtractNode()|, \verb|ExtractName()|, and \verb|ExtractInterfaceDetails()| parse specific details from textual descriptions. \verb|AssignUniqueID()| assigns unique network identifiers, ensuring each component is distinctly recognized in the simulation environment. Together, these elements systematically transform plain text input into a structured format that is both accurate and suitable for any layer3 topology generation.
Figure~\ref{fig: key-value pairs output} shows Key-Value pair dictionary based JSON output is a generated template for one device to show case the format of the algorithm output.


\vspace{-2mm}
\subsection{Network topology provisioning}
The integration with the simulation environment, EVE-NG, for network topology provisioning leverages the previously structured JSON containing key-value pairs of all devices, including their technical configurations. This JSON serves as the blueprint for the entire network topology within EVE-NG.\\ 
\noindent{\bf Preparation and Initialization:}
The process begins by parsing the structured JSON, which details each network device's required configuration such as device type, interfaces, and routing protocols. This data guides the creation and configuration of each virtual device within EVE-NG.\\
\noindent{\bf Node Creation:}
The \verb|create_node| function dynamically creates nodes in EVE-NG based on the specifications extracted from the JSON. It configures various attributes like device type, associated images, and hardware specifications (CPU, RAM). Depending on whether the node is a router, switch or PC, specific templates and additional parameters such as QEMU options are set. \\
\noindent{\bf Network Linking:}
Following node creation, the \verb|create_network| function establishes links between the nodes. This function reads interface details directly from the JSON and uses them to configure correct connections, ensuring that all network interfaces are linked as per the topology requirements.\\ 
\noindent{\bf Operational Execution and API Interaction:}
With nodes and networks in place, the script executes operational commands to activate and initially configure the devices via the EVE-NG API. This includes setting up interfaces and applying any predefined network routes or policies as specified in the JSON. Each action is carefully monitored through the API responses to handle exceptions and ensure successful deployment.\\
\noindent{\bf Session Management and Debugging:}
The entire process is supported by robust session management, where authentication and session cookies are handled to maintain a persistent connection with the EVE-NG API. Debugging information, such as device creation and network linkage statuses, is logged to assist in troubleshooting and validating the network setup.

\begin{figure}[t]
    \centering
    \includegraphics[width=\columnwidth]{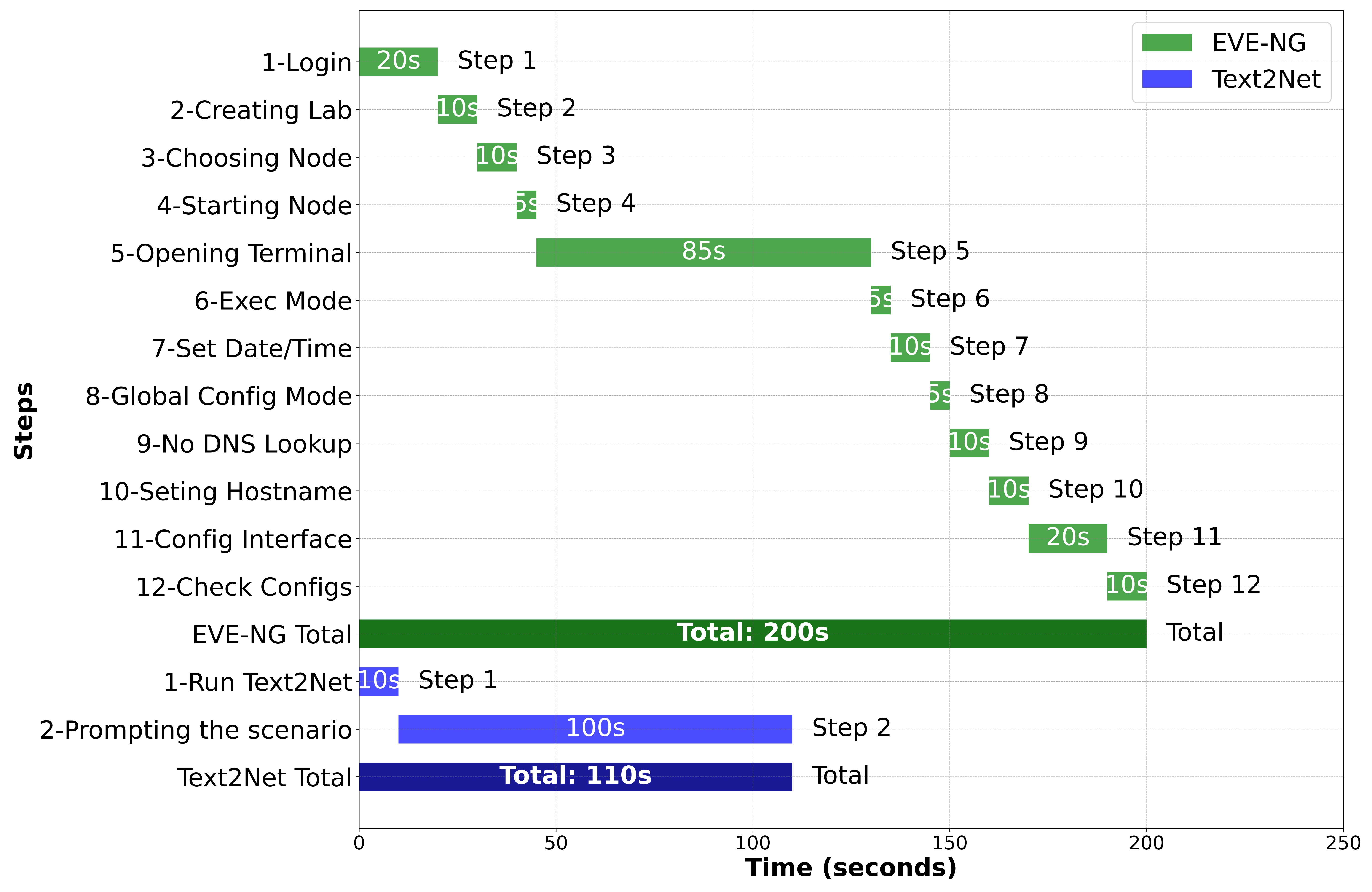}
    \caption{Steps and time comparison for Text2Net and EVE-NG Network Simulator - Scenario1}
    \vspace{-4mm}
    \label{fig: scenario1}
\end{figure}

\section{Evaluation}
The evaluation of Text2Net was conducted using both qualitative and quantitative methods. For the quantitative analysis, we compared Text2Net’s performance with manual configuration in the EVE-NG simulation environment, focusing on two parameters: time and steps. We measured only the time to input commands, excluding thinking or troubleshooting time, ensuring the results represent the best-case scenario for manual configuration. This assumption of an error-free manual process further highlights Text2Net’s competitiveness.

To assess scalability and efficiency, we analyzed three network scenarios of increasing complexity. The results demonstrate that Text2Net significantly reduces time and steps, with its advantages growing as complexity increases. This improvement stems from eliminating repetitive commands and tasks inherent in traditional workflows.

Scenario 1 involves configuring a router with basic settings, including date/time, hostname, disabling DNS lookups, configuring an interface, and verifying the configuration. As shown in Figure~\ref{fig: scenario1} (Gantt chart), manual configuration in EVE-NG requires 12 steps, such as launching the simulator, logging in, creating a lab environment, starting a node, and configuring the interface. These steps, common across network platforms, take 200 seconds.
In contrast, Text2Net completes the same task in just two steps and 110 seconds with the prompt:
\textit{“Configure a router as R1 with basic setup. Configure the interface Fast Ethernet 0/1 with IP address 192.168.0.1 and subnet mask 255.255.255.0, and finally check the configurations.”}

Scenario 2 introduces greater complexity with two routers, each having internal networks configured as loopback interfaces and interconnected with static routes. The steps from Scenario 1 are repeated for each node, including configuring loopback interfaces, setting static routes, verifying configurations, and running ping tests. Manual configuration in EVE-NG requires 510 seconds, while Text2Net completes the task in 250 seconds using the prompt:
\textit{“Configure Router 1 as R1 and Router 2 as R2 with basic configurations. On R1, configure the interface Fast Ethernet 0/1 with IP address 192.168.0.1 and subnet mask 255.255.255.0. Configure loopback 1 interface to act as Network 1 with IP address 192.168.1.1/24. On R2, configure the interface Fast Ethernet 0/1 with IP address 192.168.0.2 and subnet mask 255.255.255.0. Configure loopback 1 interface to act as Network 2 with IP address 192.168.2.1/24. Set static routes from R1 to R2 and vice versa. Finally, check the configurations.”}

\begin{figure}[t]
    \centering
    \includegraphics[width=\columnwidth]{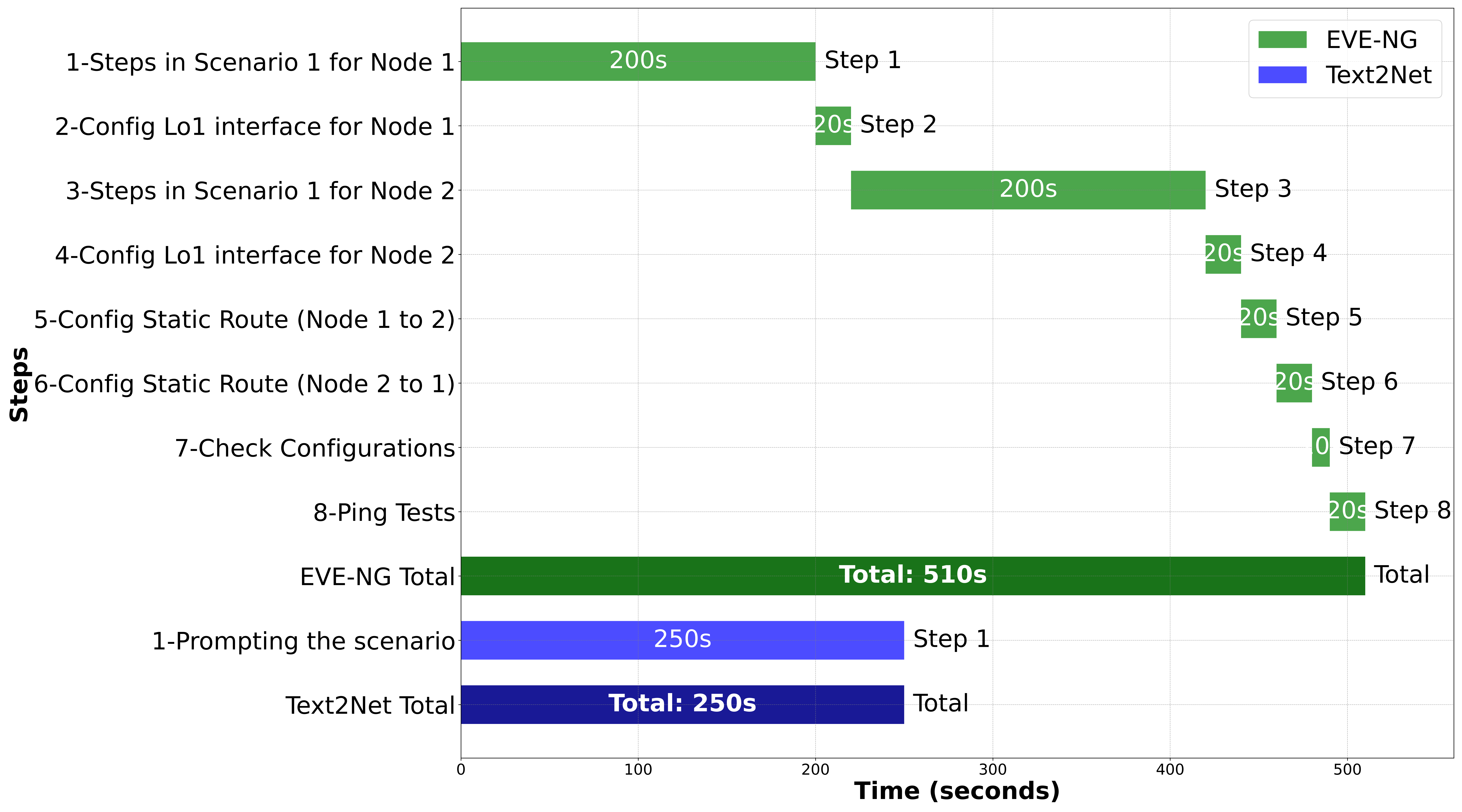}
    \caption{Steps and time comparison for Text2Net and EVE-NG Network Simulator - Scenario2}
    \vspace{-4mm}
    \label{fig: scenario2}
\end{figure}

\begin{figure}[t]
    \centering
    \includegraphics[width=\columnwidth]{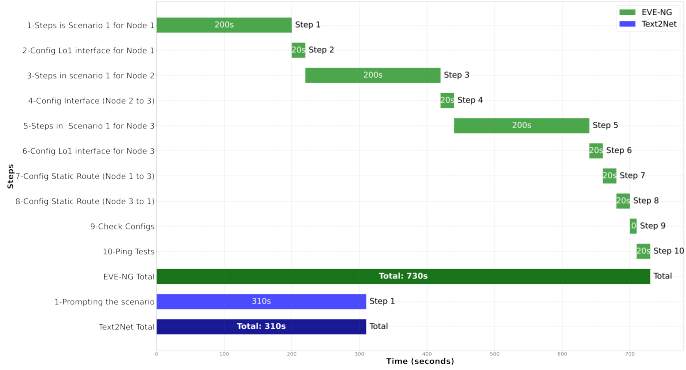}
    \caption{Steps and time comparison for Text2Net and EVE-NG Network Simulator - Scenario3}
    \vspace{-4mm}
    \label{fig: scenario3}
\end{figure}

\begin{figure*}[t]
    \centering
    \includegraphics[scale=0.44]{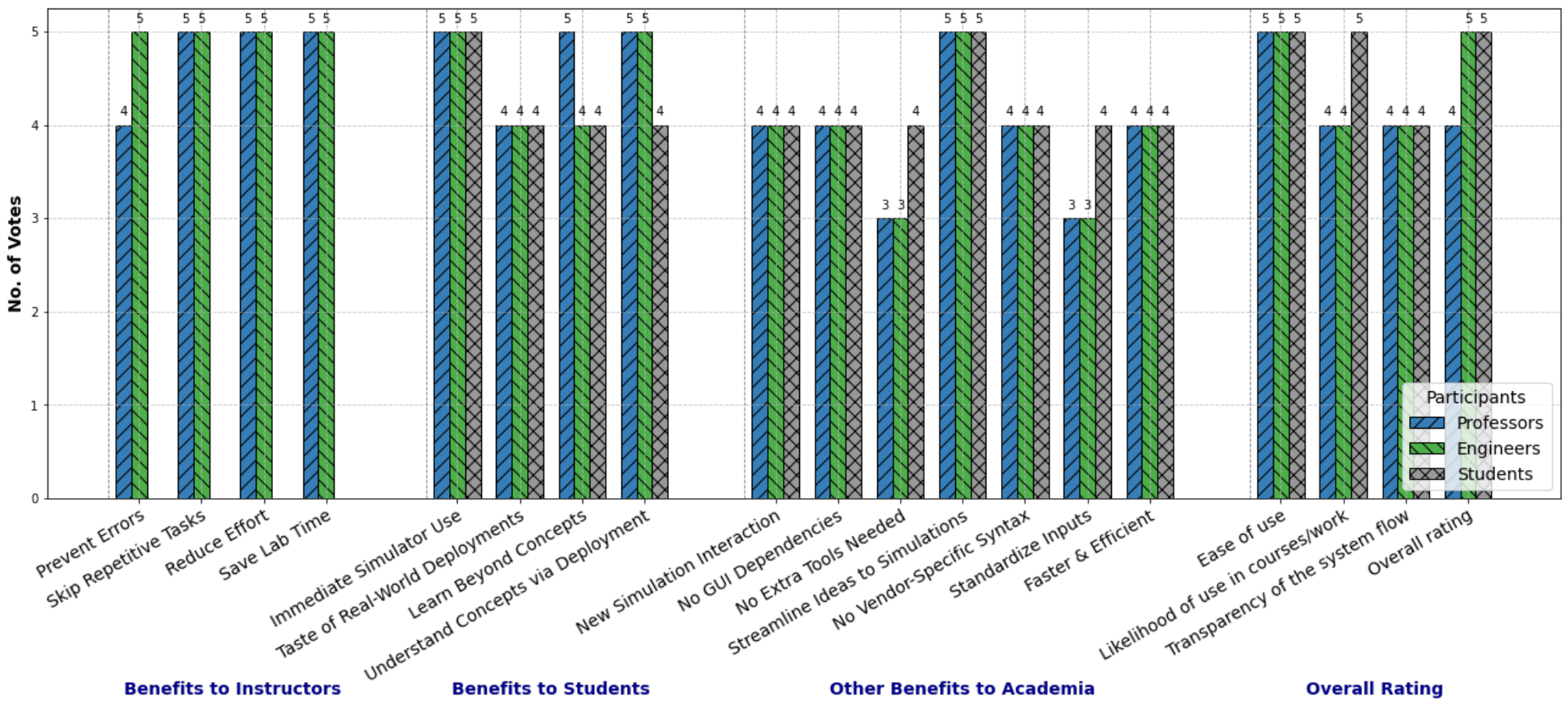}
    \caption{Consolidated Benefits and Ratings for Text2Net}
    \vspace{-5mm}
    \label{fig: consolidated_results}
\end{figure*}

Scenario 3 adds a third router, acting as a transit node between the two from Scenario 2. The static route from Router 1 (R1) now targets Router 3 (R3) via Router 2 (R2), which lacks an internal network. Manual configuration in EVE-NG requires 10 steps, including repeating all tasks from Scenario 1 for each node (R1, R2, and R3), configuring additional interface links, and setting static routes. Completing this scenario manually takes 730 seconds.
Text2Net reduces this to 310 seconds with the prompt:
\textit{“R1 is a router connected to R2. R1 interface Gigabit Ethernet 0/0 has IP address 192.168.0.1/24 and is connected to R2 interface Gigabit Ethernet 0/0 with IP address 192.168.0.2/24. R2 is connected to R3 via interface Gi 0/1 using IP address 192.168.4.1/24. R3 is connected back to R2 using interface Gi 0/0 with IP address 192.168.4.2/24. R1 has a loopback interface 1 with IP address 192.168.1.1/24 to act as internal network-1. R3 also has a loopback interface 1 with IP address 192.168.2.1/24 to act as internal network-2. A static route is configured on R1 to reach R3, and another static route on R3 to reach R1 through R2.”}

Across Scenarios 1, 2, and 3, Text2Net consistently outperforms manual configuration, requiring 110, 250, and 310 seconds, compared to 200, 510, and 730 seconds in EVE-NG. This demonstrates Text2Net’s scalability and efficiency in handling increasingly complex configurations.
Figure~\ref{fig: consolidated_results} summarizes the qualitative evaluation of Text2Net, based on feedback from 15 participants, including graduate students, professors, and engineers. Participants highlighted Text2Net’s ability to reduce errors, repetitive tasks, and setup time, making it more efficient compared to traditional methods. Additionally, the system was noted for simplifying simulation workflows and providing practical insights into real-world network scenarios. Text2Net received high ratings for ease of use, transparency, and educational value, achieving an average score of 4.66 out of 5, demonstrating its potential as a transformative tool for both academic and professional applications.

\vspace{-2mm}\section{Conclusion} 
Text2Net represents a transformative approach to network simulation, bridging the gap between plain-text user input and fully functional network configurations. By leveraging LLMs and NLP, Text2Net automates the setup of complex network scenarios, making network simulations faster, more intuitive, and accessible. The evaluation highlights its efficiency, scalability, and educational value, as it significantly reduces the cognitive and operational overhead of manual configuration. Text2Net not only accelerates the learning curve for students and educators but also demonstrates potential applications in professional settings where rapid prototyping and testing are critical. This study is limited to routing, specifically static routing, as a proof-of-concept to demonstrate Text2Net's functionality. Future work will expand its capabilities to include Layer 2 protocols (e.g., VLAN and Spanning Tree Protocol) and advanced configurations like NAT and VPN. Additionally, we plan to replace RegEX and pattern matching with LangChain and RAGs, enabling dynamic retrieval and generation of accurate network commands. These enhancements will improve scalability and establish Text2Net as a versatile tool for network simulation, education, and professional use.

\vspace{-2mm}\section{Acknowledgements}
This work was supported by the National Science Foundation (NSF CAREER CNS-2146267).
\vspace{-2mm}

\bibliographystyle{IEEEtran}
\bibliography{REFERENCES}

\end{document}